# Design and Implementation of an Emotion Analysis System Based on EEG Signals

ZHANG Yutian,   HUANG Shan,   ZHANG Jianing,   FAN Ci'en

School of Electronic Information, Wuhan University, Wuhan Hubei 430072, China

**Abstract：** Traditional brain-computer systems are complex and expensive, and emotion classification algorithms lack representations of the intrinsic relationships between different channels of electroencephalogram (EEG) signals. There is still room for improvement in accuracy. To lower the research barrier for EEG and harness the rich information embedded in multi-channel EEG, we propose and implement a simple and user-friendly brain-computer system for classifying four emotions: happiness, sorrow, sadness, and tranquility. This system utilizes the fusion of convolutional attention mechanisms and fully pre-activated residual blocks, termed Attention-Convolution-based Pre-Activated Residual Network (ACPA-ResNet).In the hardware acquisition and preprocessing phase, we employ the ADS1299 integrated chip as the analog front-end and utilize the ESP32 microcontroller for initial EEG signal processing. Data is wirelessly transmitted to a PC through UDP protocol for further preprocessing. In the emotion analysis phase, ACPA-ResNet is designed to automatically extract and learn features from EEG signals, thereby enabling accurate classification of emotional states by learning time-frequency domain characteristics. ACPA-ResNet introduces an attention mechanism on the foundation of residual networks, adaptively assigning different weights to each channel. This allows it to focus on more meaningful EEG signals in both spatial and channel dimensions while avoiding the problems of gradient dispersion and explosion associated with deep network architectures.Through testing on 16 subjects, our system demonstrates stable EEG signal acquisition and transmission. The novel network significantly enhances emotion recognition accuracy, achieving an average emotion classification accuracy of 95.1%.

**Key words：** electroencephalogram(EEG); deep learning; emotional analysis; ACPA ResNet; attention mechanism

Electroencephalogram (EEG) signals are weak electrical signals generated by neuronal activities in the brain, reflecting neural activities and playing a crucial role in fields such as medicine, neuroscience, and psychology[1][2]. Recently, EEG signals have attracted significant attention in the realm of emotion recognition research, as they not only reflect various brain activities but also effectively indicate human emotional states. However, EEG signals are low in amplitude, generally between 0.02-0.5μV when collected via non-invasive brain-computer interfaces, and their frequency and amplitude can change in response to external stimuli, thought processes, or changes in physical states[3]. Therefore, EEG acquisition devices must be highly precise. Additionally, to facilitate the collection of EEG signals, the devices should also be compact and portable.

Presently, deep neural networks (DNN) are widely applied for feature extraction and have achieved commendable results in processing images, videos, voices, and texts, outperforming traditional algorithms[4][5]. However, in EEG research, traditional emotion analysis methods still rely on manually designed feature extractors. For instance, Kumar et al. [6]used Linear Kernel Least Squares Support Vector Machines (LS-SVM) and Back Propagation Artificial Neural Networks (BP-ANN) for binary classification of emotions based on valence and arousal models, achieving accuracies of 61.17% and 64.84%, respectively. Chen et al. [7]proposed a method combining Data Space Adaptation (DSA) and Common Spatial Patterns (CSP) for EEG-based emotion classification, reaching an accuracy of 68.3% in a two-day emotion classification involving 12 subjects over five days. Compared to these, DNNs possess the ability to automatically extract and learn data features, making emotion analysis more accurate and efficient. For example, Li Youjun et al. [8] applied Stacked Denoising Autoencoder (SAE) and an LSTM-based Recurrent Neural Network (RNN) to recognize emotions from mixed physiological signals including EEG, achieving an accuracy of 79.26%. Li C et al. [9] introduced a hybrid neural network combining CNN, DNN, and LSTM, which achieved an accuracy of 75.52% on the EEG Movement/Imagery dataset. Zhou Yijun et al. [10] fused MFCC features extracted from wavelet-transformed EEG signals and utilized the features of a deep residual network (ResNet18) for emotion classification, achieving arousal and valence recognition rates of 86.01% and 85.46% in the DEAP database. Nonetheless, applying DNNs to EEG-based emotion analysis systems still poses certain challenges, and there is room for improvement in recognition accuracy, necessitating further research and optimization[11] .

In summary, current EEG signal collection and emotion analysis techniques still face deficiencies in data transmission, device portability, and accuracy of emotion analysis. Therefore, researching a real-time, portable, and highly accurate EEG signal collection and emotion analysis system is particularly crucial. Through efficient and accurate EEG signal collection and emotion classification, researchers can deepen their understanding of the relationship between emotions and EEG signals, advancing the field of emotion analysis and providing new avenues for the treatment and intervention of neurological disorders.

# 1    System Design

To address the shortcomings in EEG signal collection related to data transmission, device portability, and research costs, as well as deficiencies in emotion classification analysis, this paper aims to design and implement a simple, portable, affordable, and highly accurate emotion analysis system based on EEG signals.

The system is divided into two main parts: the front-end hardware collection and preprocessing section (Figure 1), and the back-end emotion classification algorithm section (Figure 4). Subjects generate different emotions while watching videos. The signals are collected and amplified by the analog front-end, then wirelessly transmitted via an MCU to the PC for further preprocessing and emotion classification.

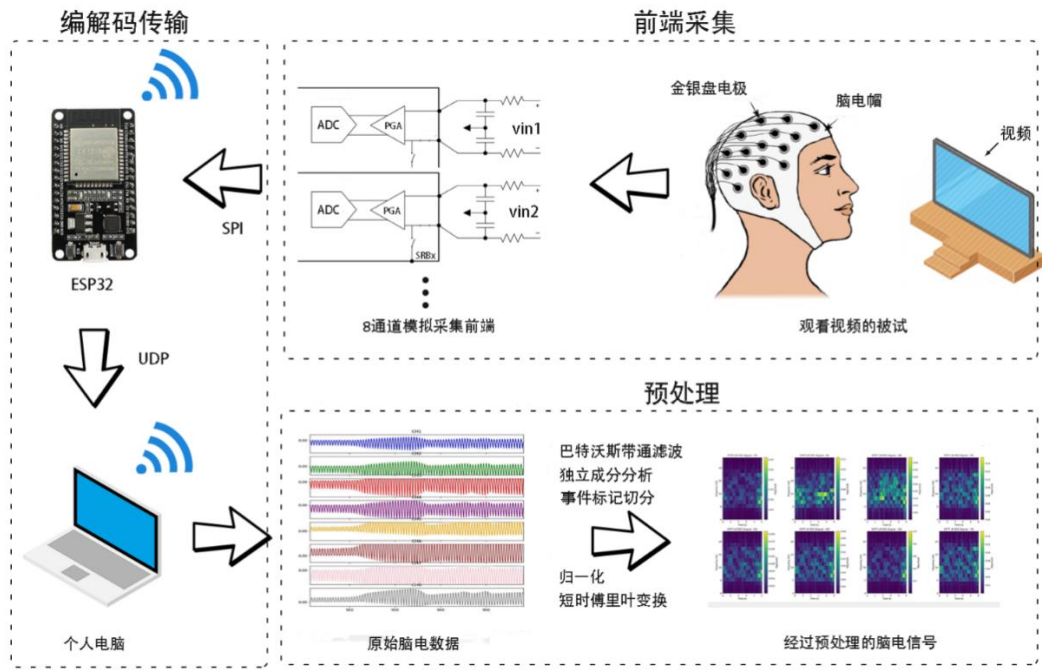

Fig.1    Frontend hardware acquisition and preprocessing process

## 1.1 Frontend Hardware Collection and Preprocessing

### 1.1.1    Frontend Collection

The collection of EEG signals relies on the accurate placement of electrodes. This system utilizes the international 10-20 system (Figure 3) as the reference standard, selecting eight electrode channels: P3, Pz, P4, O1, Oz, O2, T5, and T6.

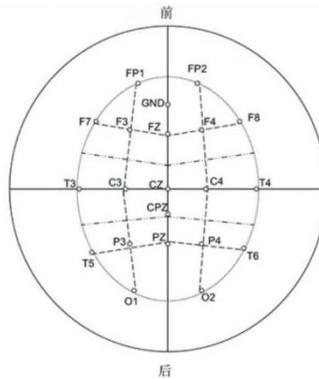

Fig.2    International 10-20 System

These electrode channels cover various areas on the posterior scalp, allowing the collected EEG signals to comprehensively reflect the subject's emotional state. Additionally, a ground electrode and a bias electrode are introduced, connected to GND and the earlobes respectively, to enhance the common mode rejection ratio and reduce interference caused by subject movement. The electrodes chosen are gold disc electrodes, and a suitable amount of conductive gel is applied between the electrodes and the subject's scalp to minimize environmental disturbances, enhance system stability, and ensure high-quality EEG signal collection.

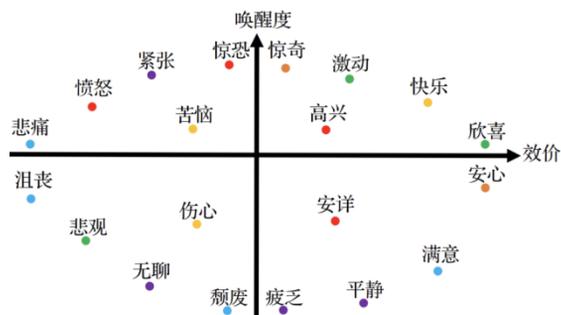

Fig.3 Two-dimensional emotion classification model

After the subjects don the collection device, they generate four types of emotions—happiness, distress, sadness, and calmness—by watching related videos and through self-reflection. The choice of emotions is guided by typical values determined by the valence-arousal (VA) model[12], a two-dimensional continuous model comprised of arousal and valence levels, as illustrated in Figure 3.

Given that the EEG signals collected via non-invasive brain-computer interfaces typically range from 0.02 to 0.5μV and are prone to disturbances, the system employs an eight-channel, 24-bit high-precision AD chip—ADS1299—as the core component of the analog frontend. After sampling, the digital signals are transmitted to the MCU via the SPI communication protocol.

### 1.1.2 Data Transmission

To enhance the system's portability, enable wireless transmission of EEG signals, and reduce research costs, the system uses a domestically produced MCU equipped with a 2.4GHz WIFI chip—ESP32—as the core component for transmitting and receiving EEG data. The MCU communicates with the analog frontend via SPI to receive eight channels of 24-bit data. These data are then decoded into eight floating-point numbers in physical units (μV), re-encoded, and transmitted in real-time to the PC listening via the UDP protocol, which is minimal in overhead and reliable. Experiments show that the data transmission delay is as low as 0.02 ms, with no packet loss observed.

### 1.1.3 Preprocessing and Feature Extraction

The raw EEG signals received by the PC are mixed with multiple frequencies and include power frequency interference and artifacts such as eye movements, making them unsuitable for direct use. The system undertakes several preprocessing and feature extraction steps. It employs an 11th-order Butterworth band-pass digital filter to extract the 5-18 Hz frequency band (α band) from the raw EEG signals. Event marking and segmentation are followed by Independent Component Analysis (ICA) to remove artifacts related to eye movements. After normalization, the short-time Fourier transform (STFT) is used to extract time-frequency domain features from the artifact-free 5-18 Hz EEG signals. The dimensionally enhanced signals are then fed into a neural network for emotion classification.

## 1.2 Emotion Classification Algorithm

After the preprocessing phase, the PC receives the feature data of EEG signals, which then enters the emotion classification phase. The system incorporates a neural network that combines convolutional attention mechanisms and fully pre-activated residual blocks, known as the Attention-Convolution-based Pre-Activated Residual Network (ACPA-ResNet), to perform emotion classification. The specific architecture of this network is shown in Figure 4.

This system architecture supports efficient processing and classification of emotional states from EEG signals, utilizing advanced signal processing techniques and neural network models to ensure high accuracy and reliability in emotion detection.

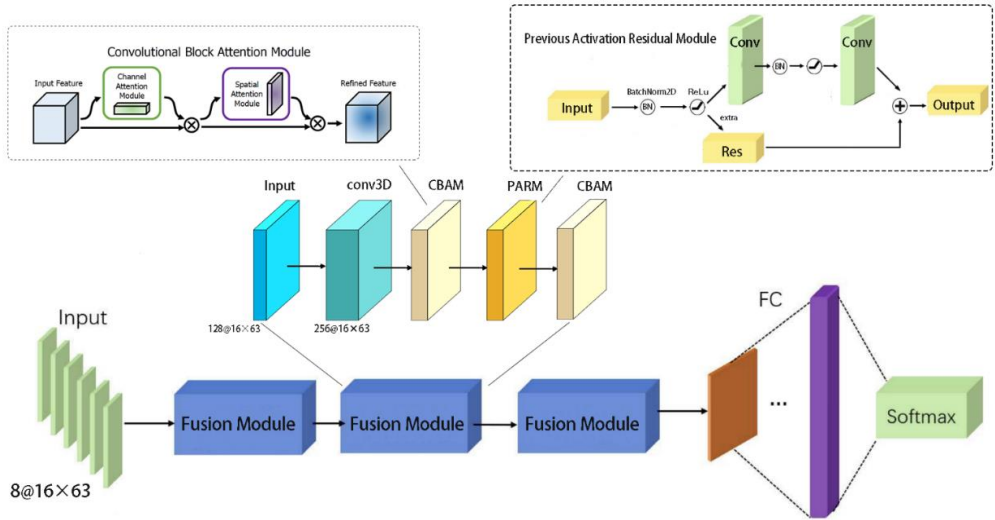

Fig.4    ACPA ResNet Network Architecture

Input Layer: This layer receives input in the form of two-dimensional features from eight channels, each channel providing a 16×63 matrix. These inputs are passed to the Fusion Module.

Hidden Layers — Fusion Module: This module consists of three parts: a convolutional layer, a Convolutional Block Attention Module (CBAM) [13], and a Fully Pre-Activated Residual Module (PARM). The convolutional layer increases the number of channels to facilitate the subsequent channel-wise attention mechanism in the CBAM. The CBAM is a simple yet effective attention module that computes attention weights in both spatial and channel dimensions, and then adapts the feature maps by multiplying them with the original feature maps, aligning well with the characteristics of EEG signals. The PARM is an improvement over the traditional Residual Block (ResBlock) [14]. Unlike the original ResBlock pattern of weight-BatchNorm-ReLU-weight-BatchNorm[15], the Pre-activation ResBlock uses BatchNorm-ReLU-weight-BatchNorm-ReLU-weight. This configuration allows gradients to flow unimpeded through fast connections to any layer in the module, thus accelerating convergence and ensuring robust performance.

Hidden Layer — Fully Connected Layer (FC): This layer uses the ReLU activation function to introduce non-linearity, enhancing the ability to capture complex patterns in the data.

Output Layer: Comprises a fully connected layer with four neurons, each outputting the prediction results for one of the four emotional states analyzed (happiness, distress, sadness, and calmness). This layer is responsible for mapping the high-level features learned by the network into the final emotion classifications.

## 2    Experiments and Results Analysis

### 2.1    Experimental Setup

The hardware setup for this experiment, as shown in Figure 5, includes an EEG cap, eight-channel gold disc electrodes, the ADS1299 acquisition module, an SPI communication adapter board, and the ESP32 module. A deep learning framework based on Python 3.7 and PyTorch 1.7.1 was established, utilizing a Lenovo Legion Y7000P with an NVIDIA GeForce RTX 2060 for training. The model was trained using ten-fold cross-validation, with 855 samples for training and the remaining 95 for testing. The average of the ten folds was taken as the recognition result for each subject. Sixteen volunteers were recruited as subjects, including individuals of different genders and emotional states. The model's recognition performance was assessed using the average recognition rate and standard deviation among these 16 subjects.

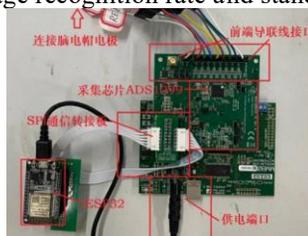

Fig.5    Physical diagram of front-end hardware acquisition system

## 2.2 EEG Signal Acquisition Results

The ADS1299 analog front-end successfully captured high-quality EEG signals from subjects. Figure 6 displays the raw EEG signals from a subject, as well as the signals after preprocessing and application of the Short Time Fourier Transform (STFT), showing clearly distinguishable EEG waveforms.

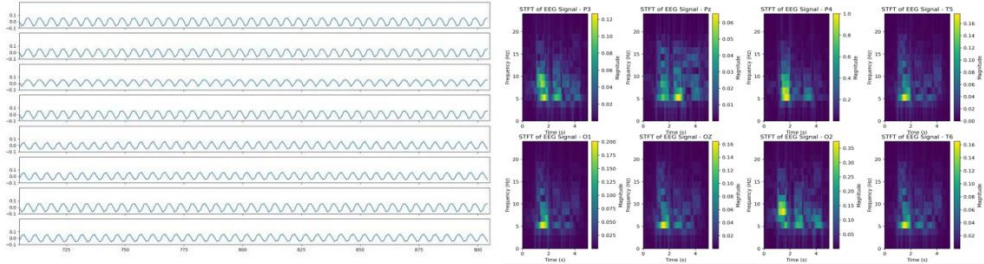

Fig.6    Original EEG signals and EEG signals passing through STFT

## 2.3 Accuracy of Emotion Analysis Algorithm

The emotion classification model, constructed using ACPA-ResNet, classified emotional states from the collected EEG signals and was compared with existing studies. System recognition accuracy was evaluated against an emotion status survey filled out by subjects during the experiments. The results, as shown in Figure 7 and Table 1, demonstrate that the system performs excellently in terms of emotion analysis accuracy, successfully classifying different emotional states.

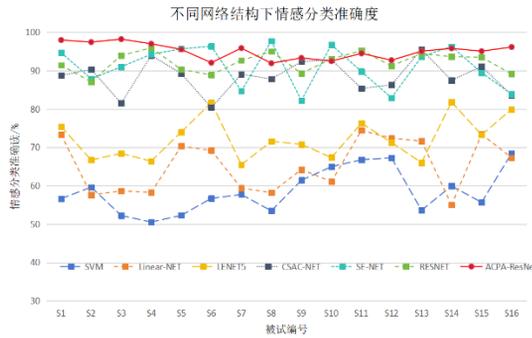

Fig.7    Accuracy of sentiment classification under different network structures

Table 1    Comparison of the accuracy of emotion recognition between this method and other methods

| 方法 | 准确度 |
| --- | --- |
| SVM | 58.6% |
| Linear-NET | 65.3% |
| LENET5 | 72.3% |
| CSAC-NET | 88.5% |
| SE-NET | 91.1% |
| RESNET | 92.2% |
| ACPA-RESNET | 95.1% |

## 3 Conclusion

This study designed and implemented an emotion analysis system based on EEG signals using the ADS1299 analog front-end and the ESP32 microcontroller. Data transmission through the UDP protocol ensured real-time and stable performance. The experiments confirmed the system's ability to obtain high-quality EEG signals and to accurately reflect the brain's activity under different emotional states. The system demonstrated excellent real-time performance and stable data transmission capabilities. It showed outstanding accuracy in emotion analysis, accurately assessing subjects' emotional states. However, there are still limitations to the system, and future improvements could focus on optimizing algorithms and hardware design, expanding the range of test subjects, and diversifying emotion state classifications. This research makes a positive contribution to the field of EEG signal collection and emotion analysis, offering new ideas and direction.